\documentclass[12pt]{iopart}
\usepackage{iopams}  

\usepackage{graphicx,xr}
\newcommand{\nvec}[1]{{\mathbf #1}}

\begin{document}

\title[ Orbital Based  Molecular Dynamics]{Design of Orbital Based  Molecular Dynamics Method}

\author{Ladislav Kocbach and Suhail Lubbad}

\address{Dept. of Physics and Technology, University of Bergen, Norway }
\ead{ladi@ift.uib.no \ \ suhail.lubbad@gmail.com}
\begin{abstract}
This paper presents a proposal of a rather new type 
of effective interatomic interaction for molecular dynamics
and similar applications. The model consists of atoms with
prescribed geometric arrangement of active orbitals, represented by 
arms of the length of half of the relevant bond lengths. The interactions 
have a repulsive part between the atomic centers and an attraction between the arm ends 
of different atoms. For each atom pair only the closest pair of arms interacts
attempting to form the bond. This is a picture of sigma bonds, the pi bonds are
modeled by alignment of additional internal vectors  which might characterize
a given atom. Also these are primarily pair interactions. Thus there are only pair
interactions of several types present. The intrinsic arrangement of the
model elements, arms and internal vectors can be switched depending on the
environment. Thus the complexity of the many-body potentials is replaced by
pair interactions between atoms with complex internal behaviour.
The proposed model thus allows formation of new geometries, 
establishing new and breaking existing bonds with the use of only
pair interactions and environment scanning.
We discuss in some detail the carbon case and shortly also hydrogen, silicon and sulfur.

\end{abstract}

\maketitle
\parindent 0cm
\parskip 0.25cm

%
%
%
%
%
      \section{Introduction}
%
%
%
%

This paper is concerned with a design of a rather new type 
of effective interatomic interaction for molecular dynamics
and similar applications. A large body of empirical data 
is concerned with so called bonded force fields, i.e. models where the structures in question are 
unbreakable but deformable, with torsion angles in addition to the stretching and compression
of the bonds. The model proposed here is of the non-bonded type of interactions,
also called reactive force fields,  which allow
formation of new geometries and establishing new and breaking existing bonds. 
%
%
During the last about 25 years several empirical 
 methods were introduced to represent basic interactions between atoms, starting with the
 three-body interactions of Stillinger and Weber \cite{Stillinger_Weber1985}, several versions of the Tersoff-Brenner potentials \cite{Tersoff_1986_Si}, 
\cite{brenner1_1990},
the so called Environment-dependent interaction potentials (EDIP) \cite{EDIP-1}, \cite{EDIP-2},  
\cite{Marks_carbon} , and the most elaborate ReaxFF \cite{REAXFF-Hydrocarbons}.
These model interactions can be used in molecular dynamics 
  without performing any quantum mechanical calculations, they are empirical representations
in some cases built on the quantum chemical studies, sometimes on models which are fitted to 
experimental facts.
The term non-bonded models could also cover
the so called {\it ab initio} methods based on quantum mechanical approach to
electronic structure, but these are generally considered separately, because 
their requirments and applications differ from the much simpler to implement 
classical potentials.

A large group of such  non-bonded interactions have been
extensively discussed in our paper \cite{kocbach_lubbad_BOP}, 
which in fact could serve as a first part of this study. 
This paper describes a proposal for a rather new approach to modeling the interactions between atoms
which is not based on a single potential function but attempts to model explicitly the 
known geometrical features of bonds and orbitals.
In simple terms it
can be described as orbital based interatomic interaction, which means that the dependence 
of the energy surface on the geometry of atomic configurations 
is modeled with
the help of predefined orbitals. We thus wish to call the proposed
model OBMD, orbital-based molecular dynamics. It will become clear that the
model leads to only pair interactions. 
We  refer  also to a recent work of
Rechtsman, Stillinger and Torquato  \cite{diamond_by_pair_isotropic}
discussing what they called
"synthetic diamond and wurtzite structures" which are self-assembled 
using only isotropic pair interactions as an example of special type of interactions.

Several of the existing potential models already contain features which are beyond
simple potential functions, as e.g. the environmental dependent interaction potentials
(EDIP) or the very detailed reactive force fields (ReaxFF). This work has been inspired 
also by these approaches and if the presented framework would 
become accepted, many empirical data could be imported from these models.

The re-orientation of the orbitals in the presented model in the MD applications would lead
to the necessity to introduce some model moments of inertia and then the presence of
spurious kinetic energy associated with the rotation of the atoms
with respect to their axes, which clearly does not have any physical meaning. Thus one part
of the model is also a prescription to avoid this problem by introducing an over-damped
rotational orientation of the orbitals, much like in the
Langevin approach, but here the Langevin equations are only applied to the
spurious energy which is removed by spurious friction. 

In the typical potential-based applications one is using a potential with predefined parameters
to simulate a physical situation. From a possible disagreement the parameters of the potential 
can be modified, and the aim is to model the chemistry with the basis of the chosen 
potential functions.
Contrary to that, the proposed model aims to include as much as possible of the 
known chemistry by giving the atoms prototypes or blueprints of the bonds
and the basic geometry  as close as possible to the chemical bonds, in a way similar to the
unbreakable bonds of molecular mechanics but with the built in reactivity feature, i.e. 
the atoms can break and form old and new bonds. At the same time the 
nature of the model is such that the atoms appear 
to interact only via two-body interactions with complex features.

The described model is not yet implemented in a full width. The parameters needed
can quite directly be derived from existing model potentials, but this should be 
followed by comparisons and readjustments to ab initio calculations.

We start by a detailed discussion 
of a simple planar model in order to introduce the idea of the OBMD in section  \ref{two-dim-sect}.
The introduction and discussion of the
overdamped angular motion is presented in section \ref{overdamped}.
Modeling of the carbon sheets in three dimensions relevant for graphene and graphite
is the subject of section \ref{3dim_sheet}.
Extensions needed
for description of carbon structures, including the possible algorithms for implementation of the models
of the structural changes are discussed in section \ref{environment}.
Possible applications of the proposed approach to other elements and compounds
are discussed in section \ref{other_compounds}.

%
%
%
%
%
      \section{Simple planar model of the $sp^2$ hybridized carbon \label{two-dim-sect}}
%
%
%
%
The interaction is modelled by overlapping orbitals belonging to each atom
(or what we imagine as hybridized atomic orbitals). 
Thus each atom looks more like a little molecule
with one "heavy" atom - i.e. the actual nucleus, 
and the centers of the orbitals, which are located at the half length of the
known bond, as a sort of additional quasiatoms. 
Here we think about the graphene sheet modelled by 
these objects with three arms. 
    \begin{figure}
    \centering
    \includegraphics[width=9cm]{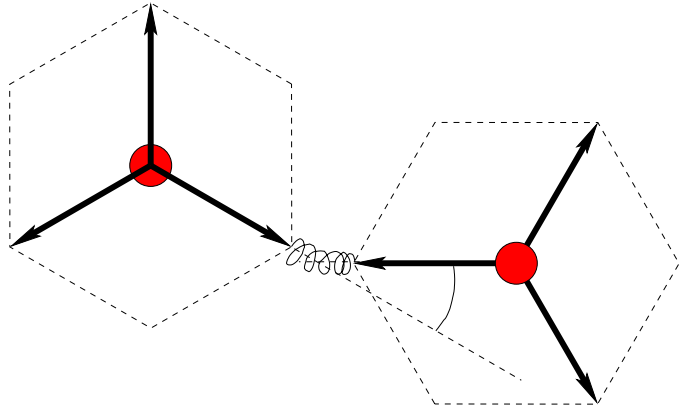}
    \caption{\small The interactions of two atoms.
    The arrows denote the vectors $\nvec{a}_{i,\alpha} $ of  eq. \ref{S_vertexangle}.}
    \label{2_atoms}
    \end{figure}
The orbitals have maximum overlap when  their centers
coincide - and when their axes are aligned,  that is the same as
when the nuclei are at 
the distance of the bond length  and their "orbital centers" coincide.

The necessary coordinates are (in 2 dimensions)
\begin{equation}
( X \ Y \ \psi )
\label{2dimcoor}
\end{equation}
The positions of the axes are rigid inside the atom, since we 
put in the model that the ideal bond length is the property
of the two neighbours. The deviation of the actual structure from 
the ideal bond is realized by  displacing the two centers and possibly
the  disalignment of the arms, the arms and vertices can only be rotated.

For characterization of N-atom system are thus needed N atomic coordinates supplemented by each atom's orientation
angle (in the plane model only)
\begin{equation}
( X_i,\ Y_i, \ \psi_i   ),\ \ \ \ \ \ \ \ i=1 . . . . N
\label{atomcoor}
\end{equation}
where we could also use a more general vector notation $\nvec{R}_i=( X_i,\ Y_i)$ for the position vectors.
    \begin{figure}
    \centering
    \includegraphics[height=6cm]{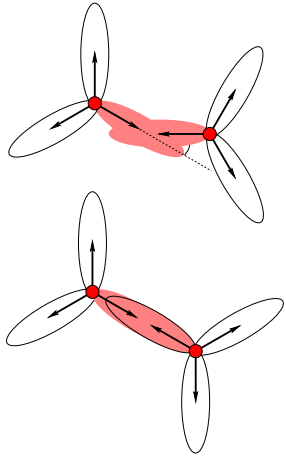}
    \caption{\small Schematic representation of the model.}    
    \label{4_atoms}
    \end{figure}
On each atom the three "arms" and vertices are characterized by their positions (vectors)
\begin{eqnarray}
x_{i,\alpha}&=& X_i + a \cos\left( \psi_i + \alpha \frac{2\pi}{3}\right) \nonumber\\
y_{i,\alpha}&=& Y_i + a \sin\left( \psi_i + \alpha \frac{2\pi}{3}\right) ,\ \ \ \ \ \ \ \ \alpha=1 . . . . 3
\label{atomparts}
\end{eqnarray}
or 
\begin{equation}
\nvec{r}_{i,\alpha} = \nvec{R}_i + a \ \nvec{a}_{i,\alpha} \ \ \ \ \ \ \ \ \alpha=1 . . . . 3
\label{atompartsvect}
\end{equation}
where  $\nvec{a}_{i,\alpha}$ are the unit vectors of the three arms. 
For each pair of atoms, their nucleus-nucleus relative distance is thus
\begin{equation}
r_{ij}  =  \sqrt{\left(X_i-X_j\right)^2 + \left(Y_i-Y_i \right)^2 }
\label{atomdist}
\end{equation}
and the nine distances between vertices eq. \ref{atomparts}
\begin{equation}
\rho_{i\alpha,j\beta}  =  \sqrt{
             \left( x_{i,\alpha}- x_{j,\beta}  \right)^2 
           + \left( y_{i,\alpha}- y_{j,\beta}   \right)^2 }
           = |  \nvec{r}_{i,\alpha} - \nvec{r}_{j,\beta}    |
\label{vertexdist}
\end{equation}
which, however, are functions of only the two atomic coordinate sets eq. \ref{atomcoor} (including the orientation angle).
The distances between the vertices of i-th and j-th atom, $\rho_{i\alpha,j\beta} $
must be accompanied by the 9 alignment parameters, or direction cosines
explicitely given by
\begin{equation}
\cos\theta_{i\alpha,j\beta}  = \frac{1}{a^2}\left[
             \left( x_{i,\alpha} - X_i \right)  \left(x_{j,\beta}- X_j  \right) 
           + \left( y_{i,\alpha} - Y_i \right)  \left(y_{j,\beta}- Y_j  \right)
           \right]  
\label{vertexangle}
\end{equation}
i.e. in shorthand notation
\begin{equation}
\cos\theta_{i\alpha,j\beta}  =  \ 
             \nvec{a}_{i,\alpha} \cdot  \nvec{a}_{j,\beta} 
\label{S_vertexangle}
\end{equation}
For each pair of atoms there is thus only two-body  potential,
strongly repulsive between the nuclei and weakly attractive
between the vertices, or orbital centers of the two atoms. That is, the nucleus-nucleus repulsion takes the form
\begin{equation}
   V_{ij}\left(  r_{ij}  \right)
               \label{V_atomdist}
\end{equation}
and the strongly coordinating nine-term potential is
\begin{equation}
   W_{ij}\left(  \nvec{R}_i,  \nvec{R}_j, \psi_i  , \psi_j \right)
   = \sum_\alpha \sum_\beta 
       w \left(  \rho_{i\alpha,j\beta},\cos\theta_{i\alpha,j\beta} \right)
               \label{W_ij}
\end{equation}
This all-vertex-vortex interaction is assumed only in this section, and it was the original idea of the
model, which could be written as a potential. 

In the present model described below, we postulate an additional saturation feature, i.e. only the
pair of nearest vortices interacts, once the nearest neighbour is established,
the remaining two on each atom are forgotten, they do not interact. It is a much more realistic model,
only one electron from each atom is forming the covalent bond (the fact that there are
so called double bonds will be discussed below).

To assure that the two vertices form a bond of the required bond length, the two arms must point
against each other, other configurations must lead to higher energy.
The simplest implementation (and thus perhaps not sufficiently flexible for all future applications) 
is a form of $w(\rho, \cos\theta)$ containing a sum of an only radial term and  one product term  
\begin{equation}
     w(\rho, \cos\theta)  =
             g(\rho) \ + \   f(\rho) \  t( \cos\theta)
               \label{w_rho_theta_prod} 
\end{equation}
where both the radial functions $g(\rho)$, $f(\rho)$ and the angular function $t( \cos\theta)$ are strongly peaked at $\rho=0$ and $ \cos\theta=-1$, respectively. 


In the discussed simple planar model
we find that the  angular alignment can be maintained by the natural stretching
resulting from the combination of the nuclear repulsion and the attraction $g(\rho)$
of the vertices,
i.e.  in 2 dimensions 
it might be enough to consider only the first term of eq. \ref{w_rho_theta_prod}
without any angular term $t( \cos\theta)$. The above addressed feature of saturation,
i.e. only one pair of vortices forms the bond, remains true also in some three dimensional
cases. Thus the atom-atom interaction eq. \ref{W_ij} which had 9 terms reduces to
one term only,
\begin{equation}
   W_{ij}\left(  \nvec{R}_i,  \nvec{R}_j, \psi_i  , \psi_j \right)
   = 
       w \left( \rho_{i\alpha_{<},j\beta_{<}},\cos\theta_{i\alpha_{<},j\beta_{<}} \right)
               \label{simpler_W_ij}
\end{equation}
where the two indices  $\alpha_{<}$ on i-th atom and  $ \beta_{<} $
 on j-th atom identify the pair of arms with the shortest distance 
 between them, i.e. for all $\alpha$, $\beta$ 
$$
\rho_{i\alpha_{<},j\beta_{<}} \le \rho_{i\alpha,j\beta} 
$$   
 %
    \begin{figure}[ht]
    \centering
    \includegraphics[width=5cm]{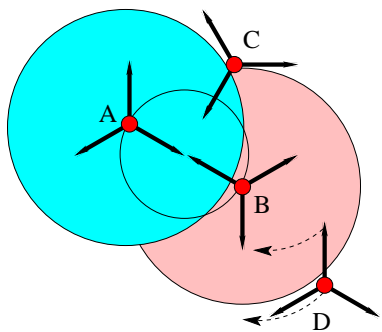}
    \caption{\small Excluding the third atom: 
    the filled areas represent the 
    repulsion areas of the potential of eq. \ref{V_atomdist}.   }
    \label{exclude_third}
    \end{figure}
Figure \ref{exclude_third}  demonstrates that only one pair of atoms
\begin{figure}[h]
\centering
    \includegraphics[width=4.5cm]{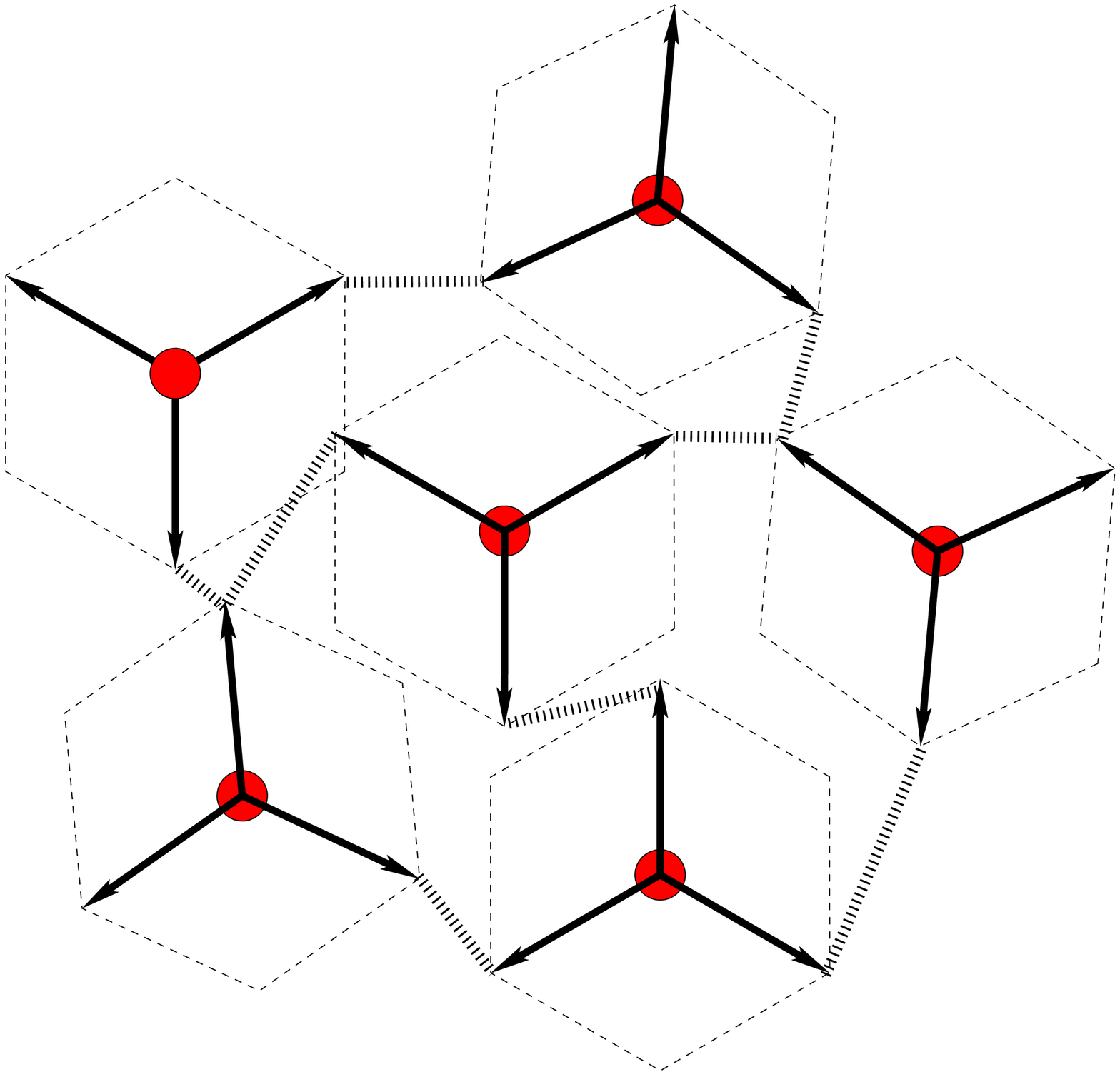}
    \includegraphics[height=4.5cm]{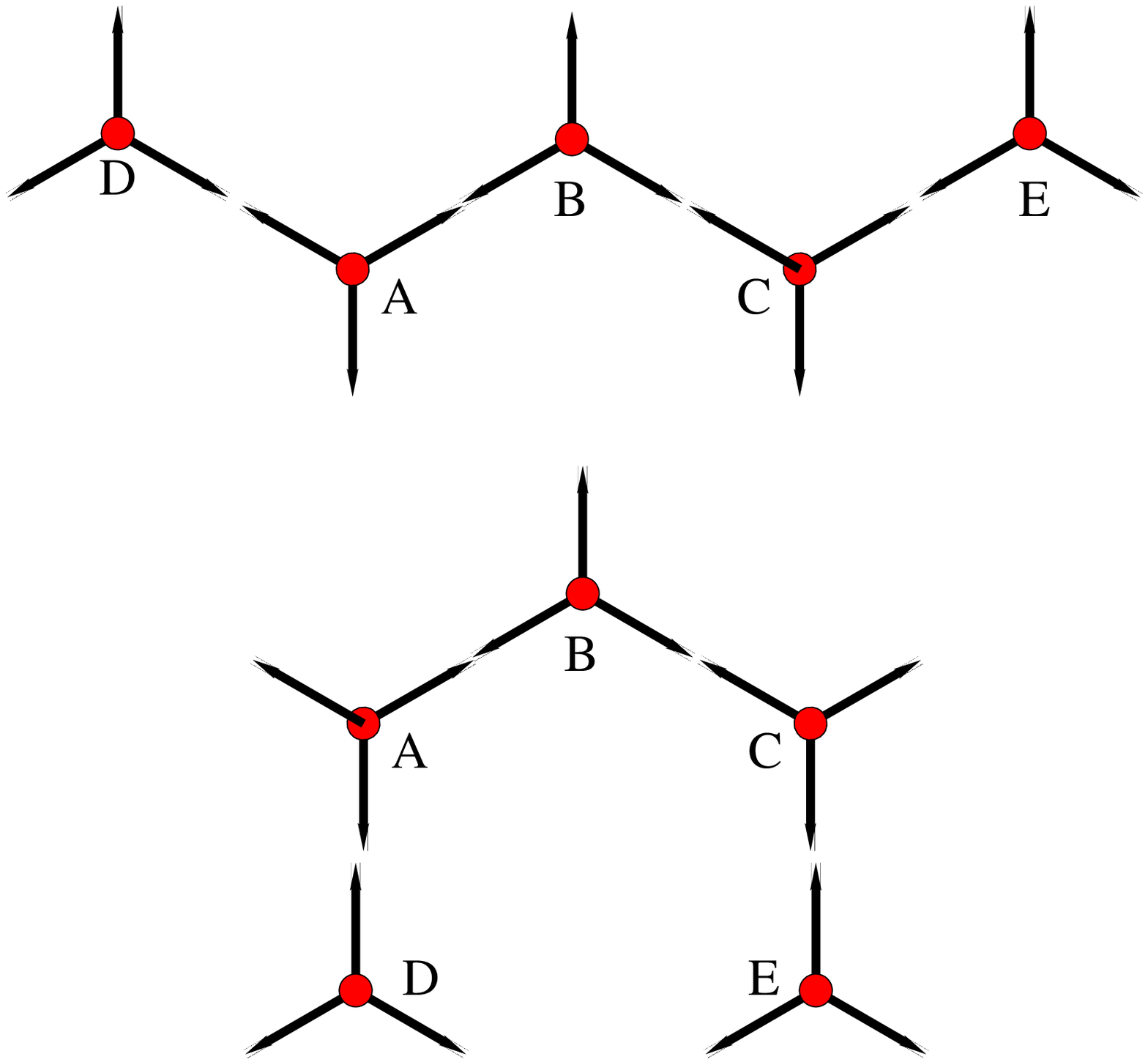}
    \caption{\small Left: The interactions of six atoms in random arrangement. 
     Right: all ways lead to graphene structure: after the first three atoms have attached themselves (A,B,C),
    the next ones can only fill in gaps (D and E). See also 
    figure \ref{exclude_third}  }
        \label{10_atoms}
    \end{figure}
can be attached. The third atom "learns" that the bond is occupied
from the over-all repulsion potential of eq. \ref{V_atomdist}.
The repulsive part and the attraction between the arms
combine together to give a potential of the same type as all
of the known interatomic potentials.
For three and more atoms the attraction of the orbitals
will be outweighed by the steep repulsion between atoms.
%
 %
%
%
%
In the planar geometry the alignment could be established simply by the
interplay of the repulsion and attraction, without considering the
angular alignment function   
$t( \cos\theta)$ of eq. 
               \ref{w_rho_theta_prod}.
 At present we leave open the question whether this 
 additional function is always necessary, but clearly its presence  
 will make the fitting of the present model
 generally possible to most of, if not all, the known interactions. 
%
%
%
%
%
\subsection{Example of a radial shape of the interatomic potential \label{two_gauss_pot} }
%
%
%
%
%
%
%
%
%
%
%
It is instructive to consider the interaction energy as a function of distance between
two nuclei with all the other parameters constant, when the "arms" are kept aligned.
This radial form of the potential should have a typical intermolecular potential shape.
\begin{figure}[ht]
   \centering  
    \includegraphics[width=7cm]{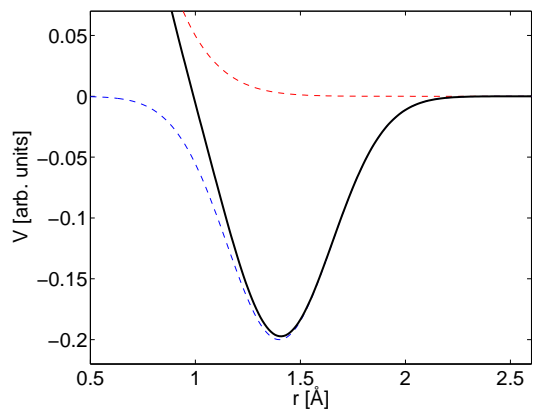}
    \caption{\small    The radial shape of aligned-arms interatomic potential described by
    eq. \ref{two-gauss-eq} with parameters discussed in the text.  }
        \label{two-gauss-fig}
    \end{figure}
This radial form is easily obtained, it is given by the sum of the expression for the 
repulsion potential and the attractive potential of the two aligned arms. If the distance 
between the nuclei is denoted $r$ and the length of each arm is $a$, then the radial form of ${\cal V} (r)$
is given by
\begin{equation}
{\cal V}(r)= V(r) + g(| r - 2 a | ) 
\label{two-gauss-eq}
\end{equation}
where $  V(r)$ is the repulsion term of eq. \ref{V_atomdist}  and $g(\rho)$ is the radial part of the
arm attraction potential of eq.  \ref{w_rho_theta_prod}.

It is quite interesting that by substituting two Gaussian shapes for both $  V(r)$ and $g(\rho)$,
we always obtain a reasonable shape for this fixed orientation interatomic potential of our model.
For example, using a=0.7 {\AA} and the following two functional shapes
$$
 V(r) = V_0  \ e^{-3 x^2}
$$
and
$$
 g(\rho) = - 0.2 V_0  \ e^{-8 \rho^2}
$$
we obtain the potential shown in figure \ref{two-gauss-fig}.
%
%
%
%
%
      \section{Orientation of the bonds: overdamped motion model \label{overdamped}}
%
%
%
%
The equations to be obtained from Lagrange function. Schematically:
%
%
\begin{equation}
{\cal I}\ddot{\psi}= -A \dot{\psi} - \frac{\partial V}{\partial \psi}
\end{equation}
where ${\cal I}$ is a fictious moment of inertia of the atom and $A$ is also
fictious coefficient of "angular friction". These two parameters would be needed
in  the model to assure a reasonable behaviour of the simulations.
A method which avoids this additional complexity is to assume that
the angular motion is "overdamped", anologous to the over-damped Langevin motion. This means that  
$$
-A \dot{\psi} \  \gg  \  {\cal I}\ddot{\psi}
$$
where $A$ must be adjusted (as a free parameter)
and instead of the second order equation the angles are updated according to
\begin{equation}
 \dot{\psi} = -\frac{1}{A}   \frac{\partial V}{\partial \psi}
\end{equation}
%
%
%
%
%
%
%
      \section{The graphene sheet model in three dimensions   \label{3dim_sheet} }
%
%
%
%
In this section we discuss the extension of the previous model to 3 dimensions.
The model is extended by the rotation   axis of each atom, perpendicular to the
plane where $\psi $ is measured. The unit vector is given by the two customary angles
$\theta$ and $\phi$. The model thus consists of atoms which contain three arms 
as in previous section and in addition a specified direction vector $\nvec{n}$ which 
is perpendicular to the plane of the arms. The variables describing the atom are
thus the arms as in previous section and the additional direction vector, specified by
the two angles (analogy to the Euler angles). 

For characterization of N-atom system are thus needed N atomic coordinates 
$\nvec{R}$ supplemented by each atom's three orientation
angles 
$\theta$, $\varphi$ and $\psi$ 
\begin{equation} 
( \nvec{R}_i, \ \theta_i ,  \varphi_i,  \psi_i   ),\ \ \ \ \ \ \ \ i=1 . . . . N
\label{atomcoor-3}
\end{equation}
where  $\psi$ as before specifies the rotation
of the arms while the angles $\theta$ and $\varphi$ specify the normal to the plane 
of the arms, as well as the rotational axis for the rotation by  $\psi$. Note that
we adopt a notation where
$$
 \nvec{n}_i = (\sin \theta_i \cos \varphi_i,   
             \sin \theta_i \sin \varphi_i, 
             \cos \theta_i  )
$$
The interaction between the arms is
identical to the previous section, a new element added is the simulation of
the $\pi$-bonding by pairwise alignment of the axes of the interacting atoms. 

The nature  of the $\pi$-bonding does not require the saturation feature discussed
above, rather the requirement that the two atoms engaging in this new interaction 
also already interact via one of the arms-pairs. There are now three 
angular variables instead of one and the equations for angular motion become 
more complex. The generalized gradient and torque will update the angles again 
through first order equations, since the model assumption of over-damped motion 
is also present here.

Formally, the interaction can now be written as follows.
The atom-atom interaction analogous to eq. \ref{W_ij} becomes now
\begin{equation}
   W^p_{ij}
   \left(  \nvec{R}_i, \nvec{n}_i, \psi_i  ,
                      \nvec{R}_j, \nvec{n}_j, \psi_j \right)
   = 
       w \left( \rho_{i\alpha_{<},j\beta_{<}},\cos\theta_{i\alpha_{<},j\beta_{<}} \right) 
       + g\left(  
       \left| \nvec{n}^\perp_i  \cdot  \nvec{n}^\perp_j     \right| - 1  
        \right) \ \ \ \ \
               \label{3dim_W_ij}
\end{equation}
where the two indices  $\alpha_{<}$ on i-th atom and  $ \beta_{<} $
 on j-th atom are given by the minimal distance between them 
 ( the pair out of the nine with the shortest distance),
$$
\rho_{i\alpha_{<},j\beta_{<}} \le \rho_{i\alpha,j\beta} 
$$  
for all $\alpha$, $\beta$ and the $\theta_{i\alpha_{<},j\beta_{<}}$ is the angle
between these two arms - this time in the full three-dimensional space.   
The vectors $\nvec{n}^\perp_i$ and  $\nvec{n}^\perp_j$ are the components of the vectors
 $\nvec{n}_i$ and  $\nvec{n}_j$, perpendicular to the line connecting the two
 atoms, 
 $$
 \nvec{n}^\perp_i \cdot \nvec{r}_{ij}=0    \ \ \ \ \ \ \ \  \nvec{n}^\perp_j \cdot \nvec{r}_{ij}=0 
  $$

It should be noted that the term
$$
 g\left(  
       \left| \nvec{n}^\perp_i  \cdot  \nvec{n}^\perp_j     \right| - 1  
        \right)
$$
is a new model effective interaction which
can be understood as related to a part of the dipole-dipole interaction (cf. \cite{kocbach_lubbad_DIPOLE}),
$$ 
   \nvec{n}_1^\perp \cdot \nvec{n}_2^\perp
   - 2   \nvec{n}_1^\|    \cdot \nvec{n}_2^\|
   $$
\begin{figure}
    \centering
    \includegraphics[width=5.7cm]{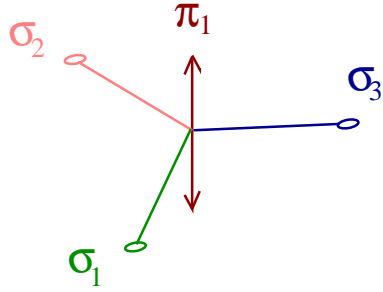}
    \caption{\small  Representation of the so called sp$^2$-hybridized orbitals of a carbon atom  } 
    \label{3-sigma-pi.png}
    \end{figure}
namely the part which would describe the attraction of two small magnets
when they are oriented antiparallel to each other and perpendicular to their connecting 
line. We discuss this particular feature and this model 
mechanism of sheet alignment in more detail in a short paper \cite{kocbach_lubbad_TDDI}
where also results of a small simulation demonstrate that this model interaction
really leads to planar alignment.
We add that the neglected part of the dipole-dipole expression above, i.e. for the 
\begin{figure}
    \centering
    \includegraphics[width=9cm]{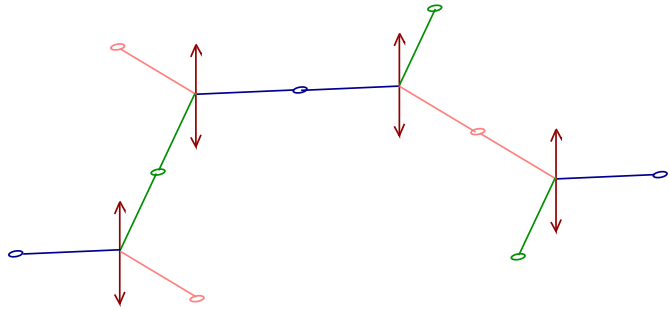}
    \caption{\small  Schematic drawing of a   group of carbon atoms with sp$^2$-hybridized orbitals forming a part 
    of the planar graphene structure. }
    \label{3-sigma-pi-plane.png}
    \end{figure}
vectors aligned with their connecting line, can be used to position 
at the desired distances of the different sheets. Thus the representation of graphite situation 
can be based on split and re-shaped dipole-dipole type interaction
where the transverse term  aligns each sheet while the parallel term has such 
radial dependence with a minimum at the known distance between sheets of graphite.
%
%
%
%
%
%
      \section{The Carbon Story  \label{environment} }
%
%
%
%
A group of carbon atoms will interact in many different ways, depending on
their mutual relations. Up to now we have discussed the situation known
as $sp^2$-hybridization of atomic orbitals. In this section we will try to 
extend this to the other known situations. The discussed models should 
describe all known carbon relations to other carbon and hydrogen atoms,
including the two diamond structures.

\subsection{Formation of carbon strings}
\begin{figure}[ht]
    \centering
    \includegraphics[width=5cm]{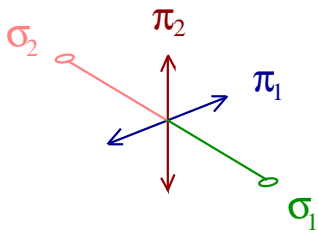}
    \caption{\small Representation of the so called sp-hybridized orbitals of a carbon atom  } 
    \label{sigma-sigma-pi-pi.png}
    \end{figure}

The free carbon atom in this model looks as shown in 
figure \ref{sigma-sigma-pi-pi.png}.
It is not really a free atom, which should have electron configuration $2s^2$    $2p^2$, but it reflects
the fact that 
a smallest stray electric field would lead to Stark states in some direction, which then would become 
the two $\sigma$  states, represented by the two arms (with centers of future
$\sigma$ bonds denoted by small circles). The two (perpendicular) $\pi$  states are
shown by the two transverse lines perpendicular to the $\sigma$  states and to each other. 

The formula for atom-atom interaction in this case is a simple modification
of eq. \ref{3dim_W_ij}  
\begin{equation}
   W^s_{ij}
   \left(  \nvec{R}_i, \nvec{n}_i, \psi_i  ,
                      \nvec{R}_j, \nvec{n}_j, \psi_j \right)
   = 
       w \left( \rho_{i\alpha_{<},j\beta_{<}},\cos\theta_{i\alpha_{<},j\beta_{<}} \right) 
       + \sum_{\mu \nu} g\left(  
       \left| \nvec{n}^\perp_{i\mu}  \cdot  \nvec{n}^\perp_{j\nu}     \right| - 1  
        \right)  \ \ \ \ \
               \label{W_ij_carbyne}
\end{equation}
where $\alpha_{<}$  and  $ \beta_{<} $
have the same meaning  as in  eq. \ref{3dim_W_ij},
while the sum runs over the  representatives  $\mu$, $\nu$ of the two $\pi$-orbitals
on each atom.
The emerging carbon strings are referred to as carbyne, and there have been 
some discussions about its forms with alternating single and triple bonds in
contrast to equivalent double bonds in a chain. It should be noted that 
the presented model can not distinguish the two alternatives unless some further
features would be added. 
\subsection{Graphene and graphite.}
In a situation when there are more than two close neighbours, it will become energetically advantageous to have more electron-orbital mixing than just 
one Stark mixture, which can lead to the formation of additional sigma orbitals (bonds) 
with neighbour atoms. This would switch the atom's behaviour
from $sp$ hybridization to the $sp^2$ hybridization shown in figure \ref{3-sigma-pi.png},
 with the interaction form already described by eq. \ref{3dim_W_ij}.

In figure \ref{sigma-sigma-pi-pi.png}
 is shown how the $sp^2$ hybridized states lead to formation of the planar carbon structures. 

The $\pi$-orbitals
on all atoms prefer to be aligned, this reflects the formation of $\pi$-bonds and in fact 
also the origin of the 
$\pi$-type electronic energy bands in benzene and in particular graphene. 
The correlation over many atoms 
of the $\pi$-type bands has an important energetic effect and this
could be built in a future version of the proposed model. For the time being we 
only consider the alignment of the $\pi$-vectors.

\subsection{Diamond-like structures}
\begin{figure}[ht]
    \centering
    \includegraphics[width=3.7cm]{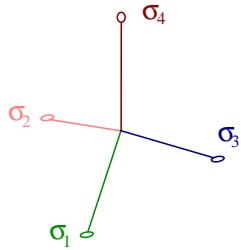}
    \caption{\small  Representation of the so called sp$^3$-hybridized orbitals of a carbon atom,
    the tetrahedral orbital arrangement}
    \label{tetrahedral.png}
    \end{figure}

When there are more than three neighbours, the energetic advantage 
leads to $sp^3$ hybridization and geometry  shown in 
figure \ref{tetrahedral.png}, i.e. there are no orbitals left in the 
pure $\pi$-form, all four orbitals lead to formation of $\sigma$-bonds.

The formula for atom-atom interaction in this case is again close to the form
of eq. \ref{3dim_W_ij}  
\begin{equation}
   W^t_{ij}
   \left(  \nvec{R}_i, \nvec{n}_i, \psi_i  ,
                      \nvec{R}_j, \nvec{n}_j, \psi_j \right)
   = 
       w \left( \rho_{i\alpha_{<},j\beta_{<}},\cos\theta_{i\alpha_{<},j\beta_{<}} \right) 
                \label{W_ij_tetra}
\end{equation}
where $\alpha_{<}$  and  $ \beta_{<} $
have the same meaning  as in  eq. \ref{3dim_W_ij}, but in this case the minimal distance
is chosen from the possible 16 instead of 9 pairs. Note that now there is no dipole-type
interaction included.

\begin{figure}[ht]
    \centering
    \includegraphics[width=6cm]{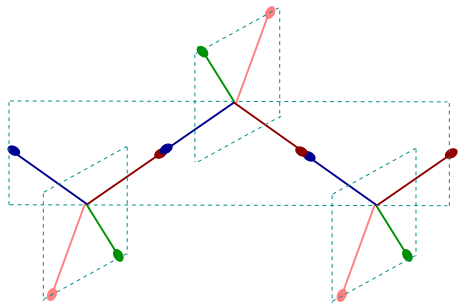}
    \caption{\small Three carbon atoms chain, occuring in diamond, lonsdaleite and in 
    aliphatic hydrocarbons. Shown is the plane defined by the three atoms, and the
    three normal planes which contain the remaining orbitals.  
    This arrangement is a property of the tetrahedral geometry}
    \label{3-carbon-aliphatic.png}
    \end{figure}
Interaction of tetrahedral $sp^3$-atoms would lead to the chain formation  of aliphatic hydrocarbons
in combination with sufficient hydrogen - or to the mesh of carbon layers  of diamond-like
and lonsdaleite-like type structures.

The basic three-atom group is sketched in figure  \ref{3-carbon-aliphatic.png}.
This three-atom group appears both in diamond, lonsdaleite and in hydrocarbons. 
The bond lengths will be different in different situations, in our model
the arm lengths are set up by each atom depending on its "knowledge"
of the neighbourhood. This shows the flexibility of the proposed approach.
One can put in various instructions for changing the atomic parameters
depending on the environment, 
i.e. every atom is a sort of automaton with instruction set allowing to 
enter as much chemistry as desirable.

\begin{figure}[ht]
    \centering
    \includegraphics[width=6cm]{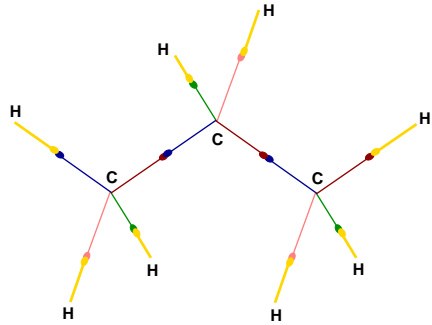}
    \caption{\small The most common isomer of the propane molecule, containing the chain of figure \ref{3-carbon-aliphatic.png}} 
    \label{propane.png}
    \end{figure}
%
%
%
%
%
%
%
%

One of the situations which can be used to adjust the interaction parameters
is represented by 
figure \ref{propane.png}
where the hydrogen nucleus is the simple vertex, a line with atom at the end, and where the
center of the sigma orbital is shown by the circle as in the other diagrams,
i.e. the molecule of propane.
\subsection{Four body correlations in diamond and lonsdaleite. \label{4-body} }
In our previous work \cite{kocbach_lubbad_BOP} we discussed how to include the four-body correlations
into the Tersoff-type and Stillinger-Weber type models.
Here we describe another feature added to the present model 
which will make it possible to energetically differentiate
diamond from lonsdaleite, i.e. build in an effective four
body correlation using the interaction form of two-body forces of the type
introduced in this paper.
\begin{figure}[ht]
    \centering
    \includegraphics[width=9cm]{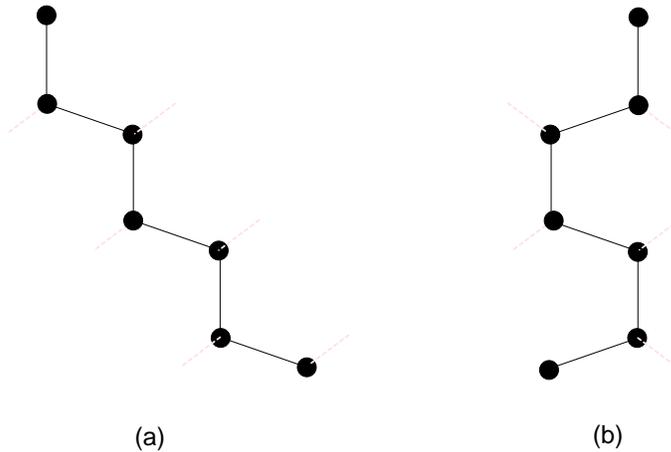}
    \caption{\small The carbon chains in diamond and lonsdaleite.
    \label{diamond-vs-lonsdale}
    }
    \end{figure}
%
%
%
%
%

The two different types of planar carbon chains present in diamond and lonsdaleite are
shown in  figure \ref{diamond-vs-lonsdale}. The chain (a) is already shown in
figures \ref{3-carbon-aliphatic.png}  and \ref{propane.png}, and it is present in both
diamond and lonsdaleite, while the chain (b) is present only in lonsdaleite.
We shall now construct an effective interaction leading to the alignment of structures
of the type (a). Comparing the two structures, it is easy to see  (especially when
using a three-dimensional molecular modeling set) that the chains of figure \ref{diamond-vs-lonsdale}
are repeated in three planes which have in common the line connecting the two atoms, i.e. one pair of the "arms".
In the case (a) the three unconnected "arms" on 
each of the neighbor atoms point in exactly opposite directions, it means that in the case (a) all four 
pairs of the arm vectors directionally closest (lying  along the same line, but in opposite sense) 
to each other are having scalar products equal to
$-1$. In case (b), and in any other conformation of carbon or aliphatic hydrocarbons, this is never the case.
The desired  effective interaction can be described by a  spin-spin type formula
\begin{equation}
U_{ij}=u(r_{ij})  \sum_{\alpha \beta } |\nvec{a}_{i\alpha}
    \cdot \nvec{a}_{j\beta} + 1 | )
\label{dipole-diamond}
\end{equation}
where $\nvec{a}_{i\alpha}$ are the unit vectors of the "arms" and
the summation runs only over the four pairs of the total 16 which 
are directionally closest, i.e. give the smallest contributions. This formula will give
zero for the diamond arrangement and a positive energy penalty for any other orientation
of the two model atoms. When additional atoms are added to the ends of the discussed "arms",
this two-body interaction results in four-body correlations.

\subsection{Environment dependent switching of interactions \label{environment_switch} }
%
%
%
%
%
%
An essential feature of the proposed model is the switching of a given atom's 
behaviour between the $sp^2$ to $sp^1$ or the $sp^3$ regimes. In the EDIP
model this is implemented by adding a more or less continuous function. 
Here we propose a sort quantum jump switching behaviour. However, it does not mean
that the structure will change by a jump. Only the relation of the arms
(and in fact their number, only the $\sigma$-orbitals are represented by arms)
If an atom can not find and get aligned
with two perpendicular $pi$-vectors, it will switch from the $sp^1$-regime 
of figure \ref{sigma-sigma-pi-pi.png} to the $sp^2$ regime of figure \ref{3-sigma-pi.png}. 

The conditions for the switching and their implementation are the strength
of the proposed model, but on the other hand they represent a problem since
at present we do not yet have a realistic prescription. This must be 
obtained both from chemical observations and from quantum chemical calculations.

When considering the $sp^3$-regime, we should mention the representation of 
hydrogen atoms, which simply are one-armed atoms with a suitable arm length
to complement the length of the carbon arm so that their sum becomes 
the length of the hydrogen to carbon bond.

%
%
%
      \section{The Silicon and Sulfur Story \label{other_compounds}}
%
%
%
%
The situation for silicon which has four valence $n=3$ electrons
is quite different from the carbon case of $n=2$ electrons.
Si-atoms would not experience the $sp^2$-hybridization of atomic orbitals
which is so important for carbon, i.e. clearly no $\pi$-bonding in addition
to the  $sp^3$-hybridization of atomic orbitals. On the other hand,
groups of Si atoms can arrange themselves in many other 
geometries (at higher temperatures). The proposed model could
be used in place of the Tersoff potential for situations where
the diamond-like structure remains the most important feature. 
For description of these cases the present model concentrating mainly
on the features seen for $n=2$ atomic states
should be extended to
model also further features of higher atomic states. 

Oxygen and sulfur are another pair of similarly related elements
with $n=2$ and $n=3$ orbitals, respectively. 
These elements show even more different behavior than carbon and silicon. 
In particular the complex behaviour of sulfur aggregates presents
a problem not well studied in the framework of simple models.
Extensions of the present model to many such new situations do not seem
to be hindered by anything else than suitable representations of further
features, derived from known chemical observations. In particular, the known
complexity of sulfur behaviour including the $S_8$ rings (cf the work of 
Stillinger and Weber on liquid sulfur \cite{liquid_sulfur_stillinger}) 
might be well modelled with the help
of future extensions to the discussed model.

%
%
%
      \section{Conclusion \label{conclusion}}
%
%
%
%

We have formulated a procedure for a design of a model empirical 
atomic interaction which has the ability to incorporate 
many more features than the traditional two- or three-body potentials. 
These many feature are directly connected with observed facts, since
the interactions are pair interactions only. This is made possible
by basing the model on predefined orbitals, much like the 
molecular structure models
invented by Andre Dreiding. A natural name for the model would thus be
the Dreiding interactions, but unfortunately the Dreiding force field already exists
and does not have any of the features discussed here.
We have thus chosen the name OBMD for the proposed approach.

The model is not yet practically implemented, we hope that we shall attract 
experts on different aspects to contribute to the fitting and modeling of the 
details. In concluding, we show how powerful the idea is by simply considering 
the possible shapes of the potentials.
In the traditional potentials, there are usually present two different inverse powers with opposite sign
(Lennard-Jones type), or two exponentials of different exponent (Morse type, Tersoff etc),  
in our approach we can combine even two Gaussians to obtain a nice shape for the potential
as shown in figure  \ref{two-gauss-fig}.
This is naturally made possible by the fact that
the two gaussians are not centered at the same point, but one at the atomic center and the
other at the orbital center.

%
%
%
%
\section*{Acknowledgments}
We would like to thank Dr Norbert L{\"u}mmen at University of Bergen for
very helpful and enlightening discussions of his work based on the ReaxFF system, 
which contributed in a very important way to the formulation of the section
on carbon.
%
%
%
%


\section*{References}

\end{document}